\documentclass[12pt]{article}
\usepackage{amsmath}
\usepackage{graphicx}
\usepackage{color}
\begin{document}
\baselineskip=18 pt
\begin{center}
{\large{\bf Relativistic quantum dynamics of spin-$0$ system of the DKP oscillator in a G\"{o}del-type space-time }}
\end{center}

\vspace{.5cm}

\begin{center}
{\bf Faizuddin Ahmed}\footnote{faizuddinahmed15@gmail.com ; faiz4U.enter@rediffmail.com}\\ 
{\bf Ajmal College of Arts and Science, Dhubri-783324, Assam, India}
\end{center}

\vspace{.5cm}

\begin{abstract}

In this article, we study the DKP equation for the oscillator in a G\"{o}del-type space-time background. We derive the final form of this equation in a flat class of G\"{o}del-type space-time and solve it analytically, and evaluate the eigenvalues and corresponding eigenfunctions, in detail.

\end{abstract}

{\it keywords:} G\"{o}del-type space-time, DKP equation, energy spectrum, wave-functions, oscillator.

\vspace{0.1cm}


\section{Introduction}

In relativistic quantum mechanics, spin-$0$ and spin-$\frac{1}{2}$ particles have been investigated via many analytical and numerical techniques such as the Nikiforov-Uvarov method (NU), supersymmetric quantum mechanics (SQM), alternative iteration method (AIM) etc.. The Duffinn-Kemmer-Petiau (DKP) equation \cite{Kemmer,Duffin,Kemmer2,Petiau} (see also \cite{HH1,HH2}) provides a good theoretical approach for describing spinless and spin-one particles. This equation is an extension of the Dirac equation \cite{PS,MH2} in which $\gamma^{\mu}$ matrices are replaced by $\beta^{\mu}$ matrices.

The Dirac oscillator were studied in \cite{DI,PAC,HU,MS} where in the free Dirac equation, the momentum operator $\vec{p} \rightarrow \vec{p}-i\,m\,\omega\,\gamma^0\,\vec{r}$, with $\vec {r}$ being the position vector, $m$ the mass of the particle, $\gamma^0$, $(\gamma^0)^2=1$ is the Dirac matrix, and $\omega$ being the frequency of oscillator.  Similarly, the DKP oscillator system is constructed by replacing $\vec{p} \rightarrow \vec{p}-i\,m\,\omega\,\eta^0\,\vec{r}$, where the additional term is linear in $\vec{r}$, $\eta=2\,(\beta^0)^2-{\bf I}$ with $\eta^2={\bf I}$ where, ${\bf I}$ is the $5\times 5$ unit matrix. A new oscillator model with different form of non-minimal substitution within the framework of DKP equation, were presented in \cite{Kuli}. Bound state energy eigenvalues for the relativistic DKP oscillator by using an exact quantization rule investigated in \cite{Kasri}. Recently, there has been a growing interest on the so called DKP oscillator \cite{AB,AB2,HH,ND,YN,YN2,YN3,AB3,IB,FY,MF,MF2,MF3,GG,MH,LBC, Hassan} with or without potentials. 

For charge-free particle of spin-$0$ system, the well-known first order relativistic DKP equation is \cite{Kemmer,Duffin, Kemmer2,Petiau}
\begin{equation}
(i\,\beta^{\mu}\,\partial_{\mu}-m)\,\Psi=0\quad (\hbar=1=c),
\label{1}
\end{equation}
where $m$ is mass of the particle, and $\beta^{j} (j=0,1,2,3)$ are the DKP matrices satisfying the DKP algebra \cite{FF}.

In this work, DKP oscillator in a topologically trivial flat class of G\"{o}del-type space-time, is study. This paper comprises as follows: in section {\it 2}, spin-$0$ system of the DKP oscillator in a flat class of G\"{o}del-type space-time is study and obtain the eigenvalues, and conclusions in section {\it 3}.

\section{DKP oscillator in a G\"{o}del-type space-time}

Consider the following stationary space-time \cite{Faiz} (see \cite{EPJC,EPJC2,EPJC3,FF})
\begin{eqnarray}
ds^2=-(dt+\alpha_0\,x\,dy)^2+h_{ij}\,dx^i\,dx^j\quad (i,j=1,2,3),
\label{4}
\end{eqnarray}
where $\alpha_0>0$ is a real number and $h_{ij}$ is the spatial part of the Minkowski metric. The properties of the above space-time have been studied, in idetail, in \cite{Faiz}. The different classes of G\"{o}del-type geometries have been studied in \cite{EPJC3}, and shown that the above space-time belongs to a linear or flat class of G\"{o}del-type space-time. The determinant of the above metric tensor is $det\,(g_{\mu\nu})=-1$.

We have defined the tetrad basis $e^{\mu}_{(a)}$, $e^{(a)}_{\mu}$ for the space-time in \cite{FF} which must satisfy the following relation 
\begin{equation}
g_{\mu\nu}=e^{a}_{\mu} ({\bf x})\,e^{b}_{\nu} ({\bf x})\,\eta_{a\,b},
\label{9}
\end{equation}
where $\eta_{a\,b}=\mbox{diag} (-1,1,1,1)$ is the Minkowski metric tensor.

For the DKP equation in curved space, partial derivatives $\partial_{\mu}$ is replaced by covariant derivatives $\nabla_{\mu}=\partial_{\mu}+\Gamma_{\mu}$, where $\Gamma_{\mu}$ is called the spinorial affine connection. Therefore, the DKP equation in curved space in the background curvature is given by
\begin{equation}
[i\,\beta^{\mu} ({\bf x})\,(\partial_{\mu}+\Gamma_{\mu} ({\bf x}))-m-\xi\,R]\,\Psi=0,
\label{10}
\end{equation}
where $\xi$ is the non-minimal coupling constant, $R$ is the scalar curvature. For the space-time (\ref{4}), the Ricci scalar curvature is given $R=2\,\Omega^2$.

The DKP oscillator of the DKP field $\Psi$ of mass $m$ in curved space-time is described by
\begin{equation}
[i\,\beta^{\mu} ({\bf x})\,(\partial_{\mu}+\Gamma_{\mu} ({\bf x})+m\,\omega\,X_{\mu}\,\eta^{0})-m-\xi\,R]\,\Psi=0,
\label{11}
\end{equation}
where
\begin{equation}
X_{\mu}=(0, x, 0, 0),\quad \eta^0=2\,(\beta^0)^2-{\bf I}.
\label{12}
\end{equation}

In \cite{FF}, we have constructed beta matrices $\beta^{\mu} (\bf x)$ and the spin-connections $\Gamma^{\mu} (\bf x)$. Since, the metric (\ref{4}) is independent of $t, y, z$. Suppose, the general wave-function to be
\begin{equation}
\Psi (t,x,y,z)=e^{i\,(-E\,t+l\,y+k\,z)}\,\left (\begin{array}{c}
\psi_{1} (x)\\
\psi_{2} (x)\\
\psi_{3} (x)\\
\psi_{4} (x)\\
\psi_{5} (x)
\end{array} \right),
\label{13}
\end{equation}
where $E$ is the total energy, and $l, k$ are constants. Substituting (\ref{13}) into the Eq. (\ref{11}), we arrive at the following differential equations :
\begin{eqnarray}
\label{14}
&&E\,\psi_{2}-i\,\psi'_{3}+l\,(\alpha_0\,x\,\psi_{2}+\psi_{4})+k\,\psi_{5}-i\,(a-m\,\omega)\,x\,\psi_3=(m+2\,\xi\,\Omega^2)\,\psi_{1},\quad\quad\\
\label{15}
&&(E+\alpha_0\, l\, x)\psi_{1}=(m+2\,\xi\,\Omega^2)\,\psi_{2},\\
\label{16}
&&i\,\psi'_{1}+i\,m\,\omega\,x\,\psi_1=(m+2\,\xi\,\Omega^2)\,\psi_{3},\\
\label{17}
&&-l\,\psi_{1}=(m+2\,\xi\,\Omega^2)\,\psi_{4},\\
\label{18}
&&-k\,\psi_{1}=(m+2\,\xi\,\Omega^2)\,\psi_{5},
\end{eqnarray}
in which a prime means a derivative w. r. t. $x$. These equations can be decoupled after a little algebra, which results
\begin{eqnarray}
\label{19}
&&\psi_{2}=\frac{1}{m+2\,\xi\,\Omega^2}\,(E+\alpha_0\,l\,x)\,\psi_{1},\\
\label{20}
&&\psi_{3}=\frac{i}{m+2\,\xi\,\Omega^2}\,(\psi'_{1}+m\,\omega\,x\,\psi_1),\\
\label{21}
&&\psi_{4}=-\frac{l}{m+2\,\xi\,\Omega^2}\,\psi_{1},\\
\label{22}
&&\psi_{5}=-\frac{k}{m+2\,\xi\,\Omega^2}\,\psi_{1}.
\end{eqnarray}
Substituting Eqs. (\ref{19})--(\ref{22}) into the Eq. (\ref{14}), we get
\begin{equation}
\psi''_{1}+a\,x\,\psi'_{1}+b\,x\,\psi_{1}+c\,x^2\,\psi_{1}=\lambda\,\psi_{1},
\label{23}
\end{equation}
where 
\begin{eqnarray}
&&a=\frac{3\,\alpha_{0}^2}{2}=6\,\Omega^2\nonumber\\
&&b=2\,\alpha_0\,E\,l=4\,\Omega\, E\, l,\nonumber\\
&&c=\alpha_{0}^2\,l^2+m\,\omega\,(a-m\,\omega)=4\Omega^2 \,l^2+m\,\omega\,(6\,\Omega^2-m\,\omega),\nonumber\\
&&\lambda=(m+2\,\xi\,\Omega^2)^2+l^2+k^2-E^2-m\,\omega.
\label{24}
\end{eqnarray}

Following the similar technique as done in \cite{FF}, one will arrive at the following eigenvalue equation
\begin{equation}
\Rightarrow\frac{\frac{b^2}{4\,(\frac{a^2}{4}-c)}-(\lambda+\frac{a}{2})}{\sqrt{\frac{a^2}{4}-c}}=(2\,n+1),
\label{33}
\end{equation}
with the energy eigenvalues associated with $n^{th}$ radial modes are
\begin{equation}
E_{n,l}=\pm\frac{\eta_0}{\sqrt{\eta_0^{2}+\frac{4l^2}{9\,\Omega^2}}}[3\,\Omega^2\,\{1+\eta_0\,(2\,n+1)\}+(m+2\,\xi\,\Omega^2)^2+l^2+k^2-m\,\omega]^{\frac{1}{2}},
\label{34}
\end{equation}
where 
\begin{equation}
\eta_0=\sqrt{1-\frac{4\,l^2}{9\,\Omega^2}-\frac{m\,\omega\,(6\Omega^2-m\,\omega)}{9\,\Omega^4}},
\label{35}
\end{equation}
where $n=0,1,2,3,.......$. This result shows that the discrete set of DKP energies are symmetrical about $E=0$ and this is irrespective of the sign of $l$. This fact is associated to the fact that the DKP oscillator embedded in a class of flat G\"{o}del-type space-time background does not distinguish particles from antiparticles. At this stage, due to invariance under rotation along the $z$-direction, without loss of generality we can fix $k=0$. Note that for $\omega \rightarrow 0$, the energy eigenvalues reduces to the result obtained in \cite{FF}. Equation (\ref{34}) is the relativistic energy eigenvalues of the DKP oscillator of spin-0 system in a G\"{o}del-type space-time. The corresponding wave-functions are similar to those obtained in \cite{FF}.

In \cite{LBC}, authors studied the DKP oscillator in a cosmic string space-time. The energy eigenvalues for spin-$0$ system is given by (see Eq. (57) in \cite{LBC})
\begin{equation}
E_{n,l}=\pm\,\sqrt{M^2+k^2_{z}+2\,M\,\omega\,(2\,n+\frac{|l|}{\alpha})}
\label{36}
\end{equation}
where $M$ is the mass, $\omega$ is the oscillator frequency, $\alpha$ is the string parameter, $n=0,1,2,3,........$ and $l=\pm\,1,\pm\,2,...$. For $\alpha \rightarrow 1$, cosmic string space-time reduces to the Minkowski metric in cylindrical coordinates $(t,r, \phi,z)$. Therefore, the energy eigenvalues (\ref{36}) becomes
\begin{equation}
E_{n,l}=\pm\,\sqrt{M^2+k^2_{z}+2\,M\,\omega\,(2\,n+|l|)}.
\label{37}
\end{equation}
Equation (\ref{37}) is the energy eigenvalues of the DKP oscillator for spin-$0$ system in Minkowski flat space metric in cylindrical system. 

For $\Omega \rightarrow 0$, the space-time (\ref{4}) reduces to the Minkowski metric in Cartesian coordinates $(t,x,y,z)$. In that case, the parameters $a=0=b$ and therefore from condition (\ref{33}), we have the following eigenvalues
\begin{equation}
E_{n,l}=\pm\,\sqrt{m^2+k^2+l^2+2\,m\,\omega\,n},\quad (n=0,1,2,3,......).
\label{38}
\end{equation}
Note that Eq. (\ref{37}) is the energy eigenvalues of the DKP oscillator for spin-$0$ system in Minkowski metric in Cylindrical coordinates whereas, the Eq. (\ref{38}) is the energy eigenvalues of the same system in Cartesian system.

Therefore from Eqs. (\ref{19})--(\ref{22}), we obtain
\begin{eqnarray}
\label{41}
&&\psi_{2\,n}=\frac{1}{m+2\,\xi\,\Omega^2}\,(E_{n}+2\,\Omega\,l\,x)\,\psi_{1\,n},\\
\label{42}
&&\psi_{3\,n}=\frac{i}{m+2\,\xi\,\Omega^2}\,(\psi'_{1\,n}+m\,\omega\,x\,\psi_{1\,n}),\\
\label{43}
&&\psi_{4\,n}=-\frac{l}{m+2\,\xi\,\Omega^2}\,\psi_{1\,n},\\
\label{44}
&&\psi_{5\,n}=-\frac{k}{m+2\,\xi\,\Omega^2}\,\psi_{1\,n}.
\end{eqnarray}

Let us study the eigenvalues and corresponding wave-functions one by one. We set $n=0,1,2$ and others are in the same way.
\begin{eqnarray}
\label{45}
&&(i)\quad n=0 \nonumber\\
&&E_{0}=\pm\frac{\eta_0}{\sqrt{\eta_{0}^2+\frac{4l^2}{9\Omega^2}}}\,[3\Omega^2(1+\eta_0)+(m+2\,\xi\,\Omega^2)^2+l^2+k^2-m\,\omega]^{\frac{1}{2}},\nonumber\\
&&\psi_{1\,0}=(\frac{3\,\Omega^2}{\pi})^{\frac{1}{4}}\,e^{-\frac{3\,\Omega^2\,x^2}{2}},\\
\label{46}
&&(ii)\quad n=1\nonumber\\
&&E_{1}=\pm\frac{\eta_0}{\sqrt{\eta_{0}^2+\frac{4l^2}{9\Omega^2}}}\,[3\Omega^2(1+3\eta_0)+(m+2\,\xi\,\Omega^2)^2+l^2+k^2-m\,\omega]^{\frac{1}{2}},\nonumber\\
&& \psi_{1\,1}=2(\frac{3\,\Omega^2}{4\pi})^{\frac{1}{4}}\,xe^{-\frac{3\,\Omega^2\,x^2}{2}},
\end{eqnarray}
\begin{eqnarray}
&&(iii)\quad n=2\nonumber\\
&&E_{2}=\pm\frac{\eta_0}{\sqrt{\eta_{0}^2+\frac{4l^2}{9\Omega^2}}}\,[3\Omega^2(1+5\eta_0)+(m+2\,\xi\,\Omega^2)^2+l^2+k^2-m\,\omega]^{\frac{1}{2}},\nonumber\\
&&\psi_{1\,2}=(\frac{3\,\Omega^2}{4\,\pi})^{\frac{1}{4}}\,(2\,x^2-1)\,e^{-\frac{3\,\Omega^2\,x^2}{2}}.
\label{47}
\end{eqnarray}

\section{Conclusions}

In \cite{LBC}, the quantum dynamics of scalar bosons embedded in the background of a cosmic string, were addressed. Author investigated scalar bosons described by the Duffin-Kemmer-Petiau oscillator in this background and obtained the energy eigenvalues. In \cite{FF}, we have studied the DKP equation for spin-$0$ system in a G\"{o}del-type space-time in a linear or flat cases ($\mu^2=0$), and evaluated the energy eigenvalues and eigenfunctions. 

In this work, the DKP equation for the oscillator in spin-$0$ system in a G\"{o}del-type space-time, is studied. The study space-time belongs to a class of flat G\"{o}del-type metrics, and reduces to four-dimensional Minkowski metric for zero vorticity parameter ($\Omega \rightarrow 0$). We could derived the final form of the DKP oscillator in this flat G\"{o}del-type space-time with background curvature. We have solved it analytically, and evaluated the energy eigenvalues Eqs. (\ref{34})--(\ref{35}). We have seen that both particle and antiparticle energy levels are members of the spectrum and are symmetrical about $E=0$. That fact implies that there is no channel for spontaneous boson-antiboson creation. We have shown that for $\omega \rightarrow 0$, zero oscillator frequency, the energy eigenvalues reduces to the result obtained in \cite{FF}. A special case corresponds to $\Omega \rightarrow 0$ have also addressed, and the energy eigenvalues (\ref{38}) is obtained which is found different from the result obtained in \cite{LBC} for $\alpha \rightarrow 1$ there.

In the context of quantum chromodynamics, Cosmology, gravity and particle and nuclear physics \cite{GG,BCC,RC}, the DKP equation has been examined. The results of this paper could be used in condensed matter physics and to Bose-Einstein condensates \cite{RC2,LMA}, and for neutral atoms also. It is well known that condensates can be exploited for building a coherent source of neutral atoms \cite{MOM}, which in turn can be used to study entanglement and quantum information processing \cite{DJ}. So we have presented some result which are, in addition to those obtained in \cite{LBC} present many interesting effects.

\section*{Acknowledgement} Author sincerely acknowledge the anonymous kind referee(s) for his/her valuable comments and suggestions.

\end{document}